\documentclass[aps,prb,twocolumn,superscriptaddress]{revtex4-1}

\usepackage{bm,wrapfig}

\usepackage{graphicx}
\usepackage{epsfig}
\usepackage{amsfonts}
\usepackage{amsmath}
\usepackage{amssymb}
\usepackage{color}

\newcommand\ba{\begin{eqnarray}}
\newcommand\ea{\end{eqnarray}}
\newcommand\be{\begin{equation}}
\newcommand\ee{\end{equation}}

\newcommand{\bE}{{\bf E}}

\DeclareMathAlphabet\mathbfcal{OMS}{cmsy}{b}{n}

\newcommand{\beq}{\begin{equation}}
\newcommand{\eeq}{\end{equation}}
\newcommand{\bea}{\begin{eqnarray}}
\newcommand{\eea}{\end{eqnarray}}

\def\ee{^{\rm{\bf E}}}

\begin{document}
\title{Role of time reversal symmetry and tilting in circular photogalvanic responses}
\author{Banasree Sadhukhan}
\email{banasree@kth.se}
\affiliation{ KTH Royal Institute of Technology, AlbaNova University Center, SE-10691 Stockholm, Sweden}
\affiliation{Leibniz Institute for Solid State and Materials Research IFW Dresden, Helmholtzstr. 20, 01069 Dresden, Germany}
\author{Tanay Nag}
\email{tnag@sissa.it}
\affiliation{SISSA, via Bonomea 265, 34136 Trieste, Italy}

\begin{abstract}
We study  the role of time reversal symmetry (TRS) in the circular photogalvanic (CPG) responses considering chiral Weyl semimetal (WSM) while a quantized CPG response is guaranteed by broken of both inversion symmetry (IS) and mirror symmetries. The TRS broken WSM yields one left and one right chiral Weyl nodes (WNs) while there are two left and right chiral WNs for TRS invariant WSM. We show that these features  can potentially cause the quantization of CPG response at higher values compared to the topological charge of the underlying WSM. This is further supported by the fact that Berry curvature and velocity behave differently whether the system preserves or breaks the TRS. We find the CPG  responses for TRS invariant type-II WSM to be quantized at two and four times the topological charge of the activated WNs while the chemical potential are respectively chosen in the vicinity of energies associated with left and right chiral WNs. By contrast, irrespective of the above choice of the chemical potential,  the quantization in CPG response is directly given by the topological charge of the activated WNs for TRS broken case.
Interestingly, we notice non-quantized peak in CPG response
when energies of WNs associated with opposite chiralities are close to each other as it is the case for TRS invariant type-I WSM considered here.
Moreover, we show that the tilt can significantly modify the CPG response as velocity in the tilt direction changes which enters into the CPG tensor through the Fermi distribution function. Given these exciting outcomes, 
the second order CPG response emerges as a useful indicator to characterize the  system under consideration.
Furthermore,  we investigate the  momentum resolved structure of CPG response to relate with the final results and strengthen our analysis from the perspective of the lattice models. 
\end{abstract}

\maketitle

\section{Introduction}

\par The Weyl Semimetals (WSMs) \cite{yan2017topological, weng2019lighting, hosur2013recent} have drawn a huge attention  
in recent years due to their exotic
properties that are mainly caused by the 
unusual Fermi arc surface states and chiral anomaly  \cite{RevModPhys.90.015001}. It has been found in WSMs that
non trivial band crossing occurs  at an even number of discrete points in the Brillouin zone.  These special gap closing points, protected by some crystalline symmetry, are
referred as Weyl nodes (WNs) and they carry a topological charge (referred as Chern number) which is a quantized Berry flux through Fermi surface enclosing it in momentum space \cite{RevModPhys.90.015001}. It is important to mention here  that upon breaking of either time reversal symmetry (TRS) or inversion symmetry (IS) or both of these symmetries in  Dirac semimentals, each twofold degenerate Dirac cone reduces to two isolated WNs of opposite chiralities \cite{jenkins20163d}. In particular, there exist minimum two  WNs of opposite chirality 
when the system breaks the TRS;
four WNs are noticed in general for system with
broken IS only \cite{PhysRevB.95.075133}. The conical spectrum and the point-like
Fermi surface at the WN are the signature of an untilted WSM namely, type-I WSMs. An interesting situation arises when large tilting of the Weyl cone results in a Lifshitz transition. This leads to a new class of materials called type-II WSMs, where the Fermi surface is no longer point-like \cite{PhysRevLett.115.265304}. These WSM phases have been realized experimentally in several inversion asymmetric compounds (TaAs, MoTe$_2$ , WTe$_2$) \cite{lv2015observation, xu2015discovery, jiang2017signature, li2017evidence, kimura2019optical}.

As expected, topological systems  become fertile grounds for investigating various quantum topological electromagnetic responses \cite{PhysRevB.84.075129, PhysRevLett.107.127205, PhysRevB.88.104412, Shuichi_Murakami_2007, sadhukhan2020first}.  The chiral-anomaly related negative magnetoresistance, and the quantum anomalous Hall effect are the immediate upshot of the topological nature of WSMs \cite{zyuzin2012topological, son2013chiral, ray2020tunable}. Apart from the 
electric transport, the exotic signatures associated with WSMs
show up in the thermal responses which have been studied theoretically \cite{landsteiner2014anomalous, sharma2016nernst, nag2018transport, zhang2018strong}
and experimentally \cite{hirschberger2016chiral, watzman2018dirac}. On the other hand, thanks to  distinct behavior of density of states at the Fermi level,
it has been shown that the electronic and thermal transport properties of type-II WSMs become markedly different from that of the associated with type-I WSMs \cite{fei2017nontrivial, yu2016predicted, nandy2017chiral, nag2020thermoelectric, schindler2020anisotropic}.  
In addition to the linear optical responses, the higher order optical responses, such as circular photogalvanic effect (CPGE) \cite{de2017quantized, flicker2018chiral, chan2017photocurrents, zhang2018photogalvanic, parker2019diagrammatic, xu2020comprehensive, zhang2019strong, konig2017photogalvanic, holder2020consequences}, difference frequency generation \cite{de2020difference}
are found to be very interesting for  
chiral topological crystals where the  mirror symmetry is broken in addition to inversion symmetry resulting in non-degenerate WNs. 
Topological chiral semimetals (SMs) can be realized in many multifold fermions such as, the transition metal mono-silicides MSi (M = Co, Mn, Fe, Rh)  \cite{schroter2019chiral, changdar2020electronic, ni2020giant}, double WSMs HgCr$_2$Se$_4$ and SrSi$_2$ \cite{xu2011chern, fang2012multi, huang2016new, singh2018tunable}  and triple-WSM like A(MoX)$_3$ (with $A=Rb$, $TI$; $X=Te$) \cite{liu2017predicted}.

 It is important to have non-degenerate WNs to obtain interesting  chiral transport behavior \cite{zhong2016gyrotropic,de2017quantized}. 
Very interestingly, the quantized behavior of CPG response,  which is  DC photocurrent switching  with the sense of circular polarization of the incident light,  happens to be a direct experimental probe to measure the Chern numbers in topological semimetals \cite{yao2020observation}. Very recently, a giant non-quantized photogalvanic effect have been reported in non-centrosymmetric type II Weyl semimetal TaAs-family \cite{zhang2018photogalvanic,  chan2017photocurrents} where degenerate WNs exist in  the  presence of mirror symmetry. The dipole moment of Berry curvature also leads to nonlinear Hall effect where non-quantized responses are observed \cite{sodemann2015quantum}. Moreover, interaction also leads to a non-quantized nonlinear
response \cite{rostami2020probing}.

Given the background on the higher order responses, we here 
probe the effect of TRS on the second order chiral transport namely, CPG response considering IS broken  type-I and type-II WSM. The CPG response is found to exhibit quantized response proportional to the topological charge of the WNs when the underlying untilted WSM breaks TRS, IS, and mirror symmetries. The Pauli blocking mechanism controls the behavior of CPG response where only one WN would participate in the transport and the other WN with opposite chirality remains inactive. 
Our aim is to investigate the CPG response when the underlying WSM, preserving the TRS, possesses four WNs. The questions that we would like to precisely answer are the following: is CPG response always proportional to topological charge of the underlying WSM? does the number of WNs matter?  and how can CPGE distinguishes between type-I and type-II WSMs with and without TRS? Much having explored on the non-quantized behavior of CPGE in presence of  degenerate WNs, we believe that our analysis for  the nature of quantized response in CPGE in presence of TRS happens to be the first study 
to the best of our knowledge.

In this work, we consider TRS broken and invariant type-I and type-II WSM to investigate the CPG response. We find that in general tilt can modify the CPG response as compared to the untilted case. For TRS broken WSM with two WNs shows quantized CPG response irrespective of the tilt except a few dissimilarities.  The magnitude of quatization here is proportional to the topological charge of a single WN. 
Interestingly, for TRS invariant type-II WSM with four WNs, CPG response can only become quantized while for type-I it becomes non-quantized. The magnitude of quantization depends on both the number of WNs and topological charge associated with each WNs.
 We find that the CPGE exhibits non-quantized peak instead of quantized plateau when energy gap between 
the  WNs with opposite chiralities is vanishingly small
as it is the case for TRS invariant type-I WSM.
Unlike the TRS broken case where CPG trace becomes quantized to two opposite values of same magnitude once the chemical potential is chosen close to the energies of two opposite chiral WNs, CPG trace exhibits quantization to two different values (twice and four times of topological charge) with opposite signs for TRS invariant WSM. These can be caused by the  structure of the  Berry curvature and velocity for TRS invariant WSMs that become different 
as compared to the TRS broken WSMs . Moreover, the window of quantization changes substantially depending on the activated WNs in the case of TRS invariant WSMs. We also study the momentum resolved CPG trace to further appreciate the numerical results obtained  from the lattice models.

The paper is organized as follows. In Sec.~\ref{sec2}, we describe the CPG response and introduce the TRS invariant, TRS broken lattice model. Next in  Sec.~\ref{sec3}, we discuss our numerical results, obtained from the lattice model and understand them from the perspective of the  low-energy model. Finally, in Sec.~\ref{conclusion}, we conclude with possible future direction.

\section{Formalism and Model }

\label{sec2}

\subsection{Circular photogalvanic effect (CPGE)}
\label{CPG_formalism}

The CPG injection current is a second order optical response when the system is irradiated with the circularly polarized light. It is defined as
\begin{equation}
\dfrac{d J_i}{dt} = \beta_{ij}(\omega) \left[\mathbf{E}(\omega)\times \mathbf{E}^{*}(\omega)\right]_{j},
\label{injection}
\end{equation}
where $\mathbf{E}(\omega)=\mathbf{E}^{*}(-\omega)$ is the circularly polarized electric field of frequency $\omega$, $i$ and $j$ index are the direction of current $J_i$ and circular polarized light field respectively. This optical activity is originated from the interband electronic transition.  
The tensor $\beta_{ij}$ is purely imaginary and only non-zero if IS is broken. In a chiral topological semimetals where inversion and all mirror symmetries are broken, WNs appear at different energies.  In this case the trace of $\beta_{ij}$ is quantized for a finite range of frequencies. On the other hand, 
if the system possesses at least one mirror symmetry symmetry, all the
diagonal components of $\beta_{ij}$ vanish leaving the
non-quantized CPG response from off-diagonal  component of $\beta_{ij}$ \cite{chan2017photocurrents}.
The CPG tensor $\beta$ can be written in general as \cite{de2017quantized, sipe2000second} :
\begin{eqnarray}
\label{eq:nu}
\beta_{ij}(\omega) &=& \dfrac{\pi e^3}{\hbar V}\epsilon_{jkl}  \sum_{\bm k,n,m}  \Delta f_{\bm k,nm} {\Delta v} ^i_{\bm k,nm} r^k_{\bm k,nm} r^l_{\bm k,mn} \nonumber \\
&\times& \delta(\hbar\omega - E_{\bm k,mn}),
\end{eqnarray}
where $V$ is the sample volume, $E_{\bm k,nm}=E_{\bm k,n}-E_{\bm k,m}$ and $\Delta f_{\bm k,nm}=f_{\bm k,n}-f_{\bm k,m}$ are the difference between $n$-th and $m$-th band energies and Fermi-Dirac distributions respectively, 
$\mathbf{r}_{\bm k,nm} = i \left<n|\partial_{\bm k}|m\right>$ is the off-diagonal Berry connection and ${\Delta v}^i_{\bm k,nm} = \partial_{k_i}E_{\bm k,nm}/\hbar= v_{i,n}- v_{i,m}$.

It is pertinent to discuss about the relation between the  response coefficient and  the incident applied intensity. Let's consider the electric fields in the $x-y$ plane, $\bE = |E| (1,i,0)/\sqrt{2}$. Therefore, the injection current induced in the $z$ direction is given by 
\begin{equation}
\partial_t J_z = \beta_{zz} \left[ \mathbf{E}(\omega)\times \mathbf{E}^{*}(\omega) \right]_{z} = i \beta_{zz} | E|^2 n_z
\end{equation}
with  $n_z = (0,0,1)$. The total injection current can be obtained by adding up the contributions from the three orthogonal directions: $\partial_t J_{T} = (\beta_{xx}+\beta_{yy}+\beta_{zz})i |E|^2$.
Under the reversal of polarization of the incident light i.e., $i\to -i$, the 
injection current changes its sign. Therefore, by experimentally measuring the injection current, one can directly estimate the CPG response that is encoded in the CPG tensor ${\rm Tr}[ \beta(\omega)]$.

The above CPG tensor reduces to a very tractable form for two band model where $n,m=1,2$. Following an analytical computation of CPG coefficient, one can find  the of the trace  CPG tensor $\beta_{ij}$ for a two band model is given by  
\begin{eqnarray}
&{\rm Tr}[ \beta(\omega)] &= \dfrac{ i \pi e^3}{\hbar^2 V} \sum_{\bm k} \Delta f_{\bm k,12} \partial_{k_i}E_{\bm k,12}\Omega_{i,\bm k} \delta(\hbar\omega - E_{\bm k,12}) \nonumber \\
&=& \dfrac{ i \pi e^3}{\hbar^2 V} \sum_{\bm k} \Delta f_{\bm k,12}  \Delta v_{i,12}\Omega_{i,\bm k} \delta(\hbar\omega - E_{\bm k,12})
\label{beta}
\end{eqnarray}
Here, $\Delta v_{i,12}=v_{i,1}-v_{i,2}$ is the velocity difference between valence and conduction band; $\Delta f_{\bm k,12}=f_{\bm k,1}-f_{\bm k,2}$ is Fermi distribution function between valence and conduction band. $\Omega_{i,\bm k}= i \epsilon_{ikl} \sum_{n\ne m} r^k_{\bm k,nm} r^l_{\bm k,mn} $ is the $i$-th component of Berry curvature. 
It is to be noted here that $\Delta f_{\bm k,12}$, reducing to $\pm 1$, 
plays very crucial role in order to allow the participation of the WNs for a given value of chemical potential $\mu$. This factor together with 
$\delta$-function determine the frequency dependence of the CPG response. We consider  $\omega >0$ to investigate meaningful transport properties.

Based on the linearized, un-tilted, isotropic model ${\bm k}\cdot {\bm \sigma}$ for WNs, it has been shown that the CPG trace measures the Berry flux penetrating through a surface \cite{de2017quantized}. Therefore, the topological charge $C$ of the WN, enclosed by the closed surface, results in a quantized CPG response. 
The quantization is observed in a certain frequency window which can be generically dependent on chemical potential $\mu$. Another interesting feature encoded in the $\delta$-function is that 
CPG response shows quantized response as long as $\omega$ is kept between two WN energies $E_L$ and $E_R$ i.e., $2|E'_L|<\omega < 2 |E'_R|$ with $E'_{L,R}=E_{L,R}-\mu$. 
For $\omega > 2 |E'_R|$, the other Weyl node contributes with opposite sign in the Berry flux and the quantization is generically lost.

 To complete the discussion, we here present the Berry curvature associated with the  topological WSM Hamiltonian. 
The Berry curvature of the m$^{\textrm{th}}$ band for a Bloch Hamiltonian $H(k)$,
defined as the Berry phase per
unit area in the $k$ space, is given by ~\cite{PhysRevB.74.085308}
\begin{equation}
\Omega^{m}_{a} ({\bm k})= (-1)^m \frac{1}{4|N_{\bm k}|^3} \epsilon_{a b c} N_{\bm k} 
\cdot \left( \frac{\partial N_{\bm k}}{\partial k_b} \times \frac{\partial N_{\bm k}}{\partial k_c} \right) .
\label{bc}
\end{equation}

\subsection{Lattice Hamiltonian for IS and TRS broken WSM}
\label{lattice_model1}
\begin{figure} [ht] 
\includegraphics[scale=0.135]{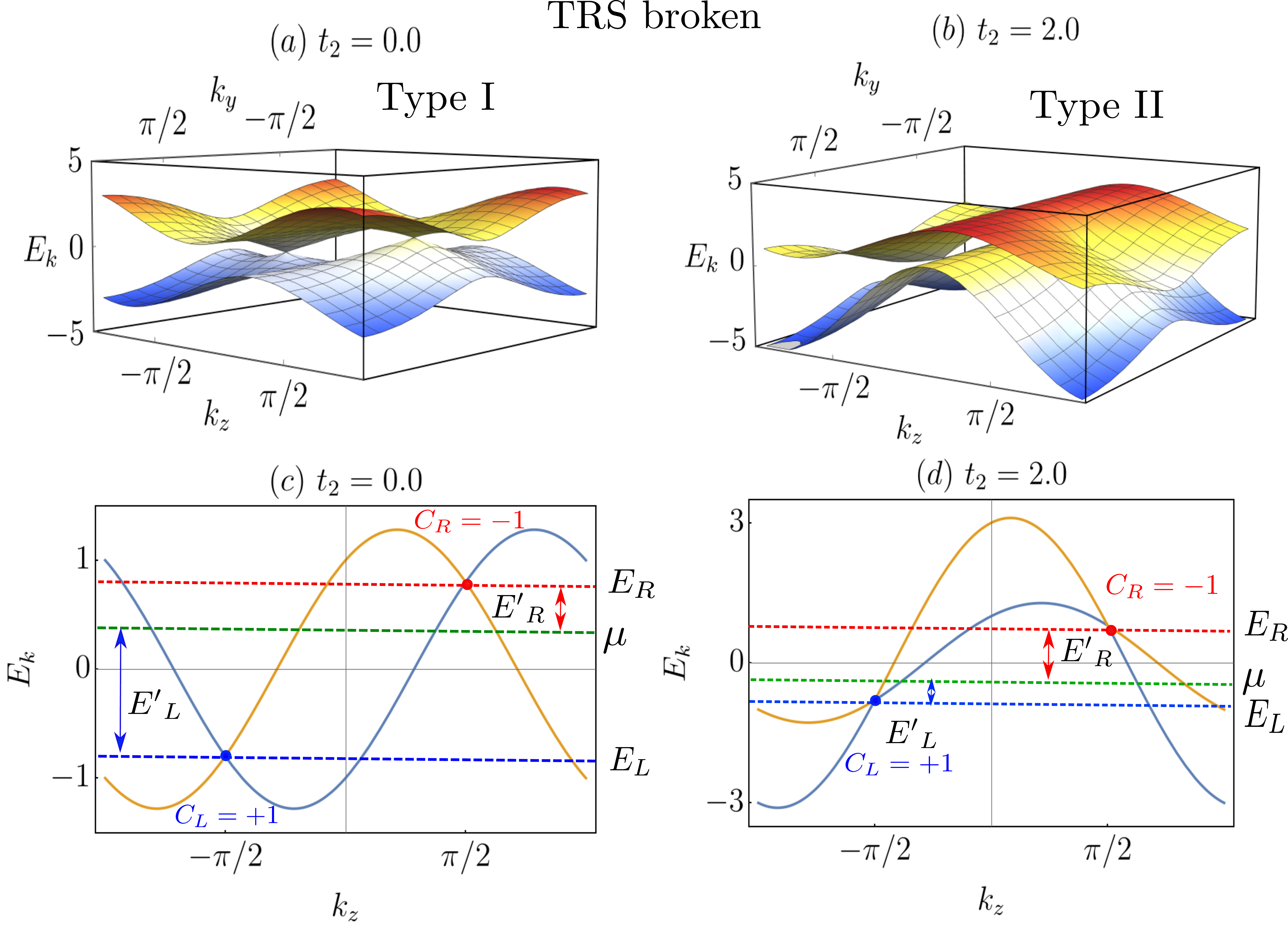} 
\caption{ The energy dispersion $E_k$
for TRS broken WSMs  are shown as function of $k_y$ and
$k_z$ for type-I (a) and type-II (b). We repeat (a) and (b) considering $k_y=0$ in (c) and (d). Two Weyl points at $k_z=\pm \pi/2$ are separated in energy $\epsilon_k=\pm \gamma$. The parameters are considered here are following: $M=2.0$, $\gamma=0.8$, $t_1=1.0$,  $t_2=0.0$ for type-I and  $t_2=2.0$ for type-II. }
\label{fig:TRS_broken_dispersion}
\end{figure}

We consider the following two band Hamiltonian for the single WSM  \cite{de2017quantized} : ${\mathcal{H}^{I}({\bm k}}) = N_{\bm k}\cdot\bm{\sigma}+ N_{0,\bm k} \sigma_0$
with
\begin{eqnarray}
N_{\bm k}&=&( t_1 \sin k_x, t_1\sin  k_y,-M+t_1\sum_{i=x,y,z}\cos k_i ) \nonumber, \\
&=& (N_{1,\bm k}, N_{2,\bm k}, N_{3,\bm k}) \nonumber \\
N_{0,\bm k} &=& \gamma \sin k_z+t_2 \cos k_z,
\label{eq:ham1}
\end{eqnarray}
where $\sigma^{0}$ is the $2\times2$ identity matrix and $\bm{\sigma}=(\sigma^{x},\sigma^{y},\sigma^{z})$ the Pauli matrices. The Hamiltonian (\ref{eq:ham1}) breaks TRS and IS: ${\mathcal T} {\mathcal H}^I(-\bm k) {\mathcal T}^{-1} \ne {\mathcal H}^I(\bm k) $ with TR operator ${\mathcal T}={\mathcal K}$ where ${\mathcal K}$ is complex conjugation; ${\mathcal P} {\mathcal H}^I(-\bm k) {\mathcal P}^{-1} \ne {\mathcal H}^I(\bm k) $ with inversion operator ${\mathcal P}=\sigma_x$. We note that Hamiltonian as represented in Eq.~(\ref{eq:ham1}) preserves $C_4$ symmetry.
The energy eigenvalues of $\mathcal{H}^{I}(\bm {k})$ are
$E_{\bm k,\pm}= N_{0,\bm k} \pm |N_{\bm k} |$ with $| N_{\bm k}|= \sqrt{ N^2_{1,\bm k} + N^2_{2,\bm k} + N^3_{1,\bm k} }$. 
For $1<\vert M/t_1\vert<3$, the model exhibits a pair of WNs of chirality $s$ at $\bm k_{s=\mp } =(0,0,\pm k_0)$ with  energies $E_{\bm k,s=\mp} = s \gamma \sin\left(k_0\right)+t_2 \cos k_0$,  where  $k_0=\cos^{-1}(M/t_1-2)$.
The right- ($s=-1$) and the left- ($s=+1$) handed  WNs now appear respectively at $E_R$ = $\gamma \sin k_0+t_2 \cos k_0$ and $E_L$ = $-\gamma \sin k_0+t_2 \cos k_0$, producing a constant chiral chemical potential $\mu_{ch}$ = $(E_R - E_L)/2$ = $\gamma \sin k_0$, which is essential to obtain a non-zero CPG response. For $t_2/t_1\to 0$ ($t_2/t_1 \to 1$), model becomes type-I (type-II) WSM. For $M=2$ and $t_1=1$, the WNs appear at ${\bm k}_{\mp}=(0, 0, \pm \pi/2)$  associated with energies $E_{R,L}=\pm \gamma$ (see Fig. \ref{fig:TRS_broken_dispersion}).

The low energy Hamiltonian close to a WN with chirality $s$ is given by
\begin{equation}
 {\mathcal H}^I_{\bm k,s} \approx s(\gamma - t_2 k_z )\sigma_0 + s k_x \sigma_x + s k_y \sigma_y + s k_z \sigma_z
\label{eq:ham1_low}
 \end{equation}
The Berry curvature takes the form $\Omega_{i}=\pm k_i/k^3$  ($i=x,y,z$) with $k=\sqrt{k_x^2 +k_y^2 + k_z^2}$. Here,  $\pm$ refers to the valence and conduction band. The velocity takes the form $v_i= \pm s k_i/k$ ($i=x,y$) and $v_z=\pm s(-t_2 + k_z/k) $. At the outset, we note that the term $ \sum_{i}^{x,y,z} \Delta v_i\Omega_i$ in CPG trace (\ref{beta})  requires separate attention for opposite chiral WNs: $ \Delta v_i\Omega_i= k_i^2/k^4$ for left chiral WN ($s=+1$) and 
$ \Delta v_i\Omega_i= - k_i^2/k^4$ for right chiral WN ($s=-1$).  
For $\mu \approx E_{L,R}$, we can consider the above low-energy model (\ref{eq:ham1_low}). 
Using the expressions (\ref{beta}) with $\Delta f=1$, we then get the CPG response as follows
\begin{align}
{\rm Tr}[ \beta(\omega)] & \approx   \frac{e^3 \pi}{\hbar^2} i \int \frac{d\Omega}{(2\pi)^3} \int k^2dk  \sum_{i}^{x,y,z} \Delta v_i\Omega_i \frac{\delta(\omega/2-E_{12}) }{2} \nonumber \\ & \approx
s\frac{e^3 \pi}{\hbar^2} i \int \frac{d\Omega}{(2\pi)^3} \int k^2dk \frac{  \sum_{i}^{x,y,z}k^2_i }{k^4} \frac{\delta(\omega/2-k) }{2} \nonumber \\ & =
i s \frac{e^3}{h^2} \oint_S d\bm S \cdot \bm \Omega=
is \frac{e^3 \pi}{h^2} C = i s\beta_0
\label{CPG_TRS}
\end{align}
Here $d \Omega$ and $d \bm S$ are the element of solid angle and surface area in a 3D geometry associated with 
 spherical polar co-ordinate.  
The above formalism clearly shows that CPG trace measures the Berry flux penetrating through $\bm S$ as discussed in Sec.~\ref{CPG_formalism}. Therefore, the topological charge $C$ of the WN, enclosed by the closed surface, results in a quantized CPG response.
Hence, from the linearized model (Eq.~\ref{eq:ham1_low}), as derived from TRS broken Hamiltonian (\ref{eq:ham1}), one can find that CPG response changes with the chirality of the WNs. This clearly suggests that  ${\rm Tr}[ \beta(\omega)]/i \beta_0$ acquires two opposite values when $\mu=E_R$ and 
$\mu=E_L$. The quantization window in terms of $\omega$ has already been discussed in Sec.~\ref{CPG_formalism}. 
We would like to comment that linearized model gives us a hint about the quantization, the lattice model however needs to be considered to get the detail of the CPG response.

We shall now address the issue of tilt in the above expression (\ref{CPG_TRS}). 
We note that the effect of tilt can only enter in the CPG response through the Fermi distribution function $\Delta f$. Interestingly,  $\Delta v_i$ and $\Omega_i$ both are tilt independent as tilt parameter $t_2$ appears in the $\sigma_0$ part of Eq.~\ref{eq:ham1_low}.  The momentum integration for tilted case would thus strongly depend on the apparently innocent factor $\Delta f $ that become $\pm 1$ for $T=0$.

For completeness, we here discuss the explicit expressions of the Berry curvature ${\bm \Omega}({\bm k})=(\Omega_x(\bm k),\Omega_y (\bm k),\Omega_z(\bm k))$ and the velocity ${\bm v}(\bm k)=(v_x(\bm k),v_y(\bm k),v_z(\bm k))$ associated with Hamiltonian (\ref{eq:ham1}) are  given by 
\begin{eqnarray}
 \Omega_x &=& \pm \frac{\cos k_y \sin k_x \sin k_z}{|N_{\bm k}|^3}\nonumber \\
 \Omega_y &=& \pm \frac{\cos k_x \sin k_y \sin k_z}{|N_{\bm k}|^3}\nonumber \\
 \Omega_z &=& \pm  \frac{-\cos k_y + \cos k_x(-1 + \cos k_y( 2-  \cos k_z))}{|N_{\bm k}|^3}\nonumber \\
 v_x &=& \pm \frac{(2-\cos k_y -\cos k_z) \sin k_x}{|N_{\bm k}|}\nonumber \\
 v_y &=& \pm \frac{(2-\cos k_x -\cos k_z) \sin k_y}{|N_{\bm k}|}\nonumber \\
v_z &=& \gamma \cos k_z -t_2 \sin k_z \nonumber \\ &\pm& \frac{(2 +\cos k_y +\cos k_z - \cos k_x) \sin  k_z }{|N_{\bm k}|}.
 \end{eqnarray}
 For a TRS broken WSM, we here find ${\bm \Omega} (\bm k)\ne -{\bm \Omega} (-\bm k)$.  Here, $\pm $ refers to the valence and conduction band.

\subsection{Lattice Hamiltonian for IS broken and TRS invariant WSM}

\begin{figure} [ht] 
\centering
\includegraphics[scale=0.14]{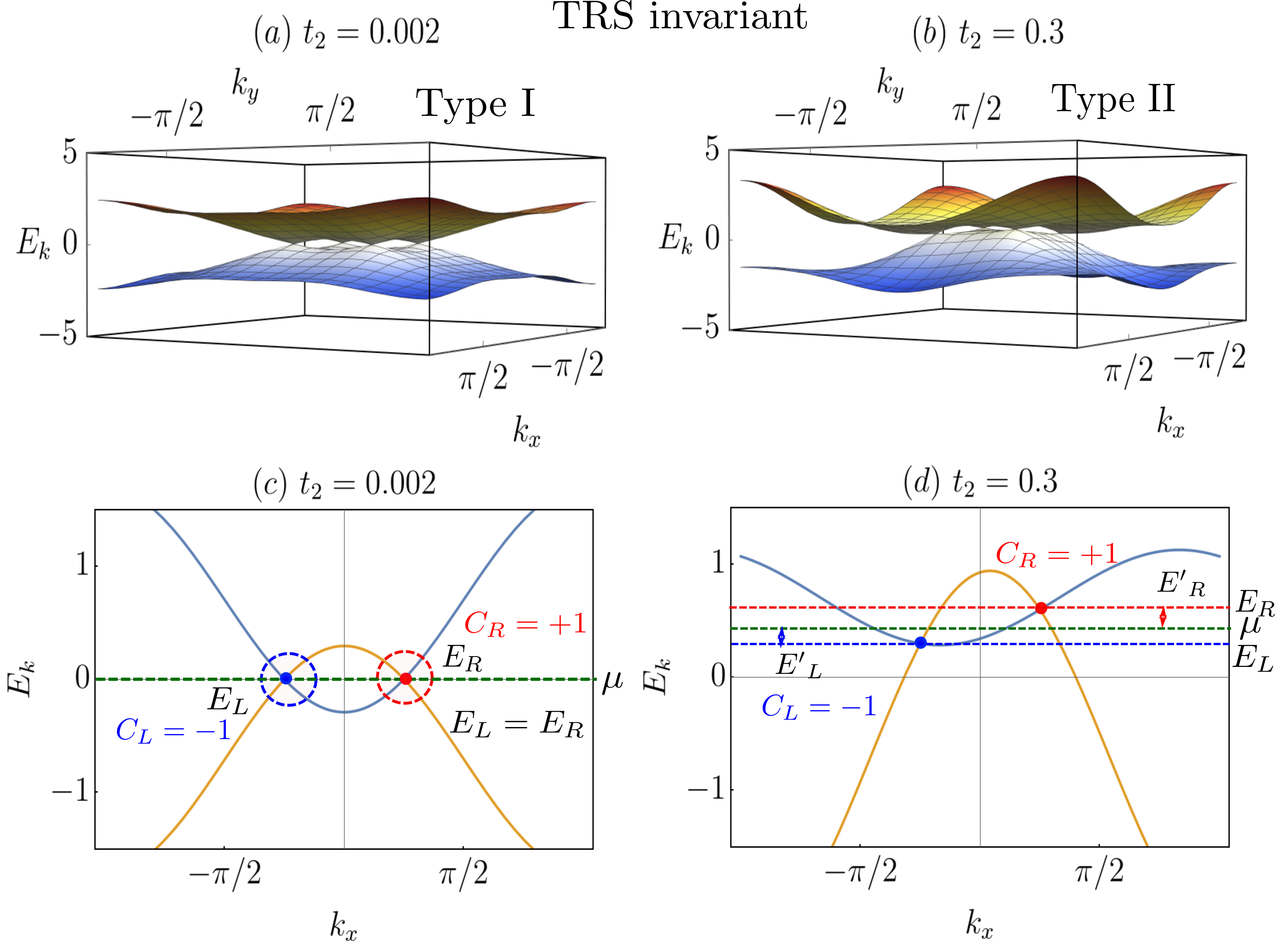} 
\caption{ The energy dispersion $E_k$
for IS broken WSMs  are shown as function of $k_y$ and
$k_z$ for type-I (a) and type-II (b). We repeat (a) and (b) considering $k_y=0$ in (c) and (d). Two Weyl points at $k_x=\pm \pi/4$ are separated in energy   $E_R-E_L=t_2 ( \delta -1)$. The parameters are considered here are following: $\delta=2.0$, $t_1=1.0$, $t_2=0.002$ for type-I and $t_2=0.3$ for type-II. 
}
\label{fig:IS_broken_dispersion}
\end{figure}

The two band  model for single WSM considered here is given by \cite{dey2020dynamic} :
${\mathcal H}^{II}({\bm k}) = N_{\bm k}\cdot\bm{\sigma}+ N_{0,\bm k} ~\sigma_0$
with
\begin{eqnarray}
N_{\bm k} &= &( t_{1} [ (\cos k_{0} - \cos k_{y}) + \delta(1 - \cos k_{z})],  \nonumber \\
&t_{1}& \sin k_{z}, t_{1} [ (\cos k_{0} - \cos k_{x}) + \delta(1 - \cos k_{z}) ] ), \nonumber \\
&=& (N_{1,\bm k}, N_{2,\bm k}, N_{3,\bm k}) \nonumber \\
N_{0,\bm k} &=& t_2 [\cos(k_x +k_y) +\delta \cos(k_x-k_y) ],
\label{eq:ham2}
\end{eqnarray}
where, $t_{1}$ and $t_{2}$ are the hopping parameters, $\delta$ ($\neq$ 1) is a constant. The Hamiltonian (\ref{eq:ham2}) breaks IS but preserves TRS: ${\mathcal T} {\mathcal H}^{II}(-\bm k) {\mathcal T}^{-1} = {\mathcal H}^{II}(\bm k) $ and ${\mathcal P} {\mathcal H}^{II}(-\bm k) {\mathcal P}^{-1} \ne {\mathcal H}^{II}(\bm k) $. It is noteworthy that Hamiltonian as represented in Eq.~(\ref{eq:ham2}) preserves $C_4 {\mathcal T} $ symmetry. 
The energy eigenvalues of $\mathcal{H}^{II}_{\bm k}$ are
$E_{\bm k,\pm}= N_{0,\bm k} \pm |N_{\bm k}|$ with $|N_{\bm k}|=\sqrt{N_{1,\bm k}^2+ N_{2,\bm k}^2+ N_{3,\bm k}^2}$. For $t_{2}$ = $0$ and $\delta >$ 1, four gapless points arise in the $k_z$ = $0$ plane and without any loss of generality we can consider $0 < k_{0} < \frac{\pi}{2}$. The right-handed ($s=+1$)
WNs are located at $ {\bm k}^{1,2}_{s=+} =\pm$ $(k_0 , k_0 , 0)$ and the left-handed ($s=-1$) WNs are located at ${\bm k}^{1,2}_{s=-}\pm$ $(k_0 , -k_0 , 0)$. When $t_2 \neq 0$,  $N_{0,\bm k}$ causes shift in energies of the WNs of opposite chiralities. The right and the left-handed WNs now appear respectively at $E_R$ = $t_2 \Big[\cos (2k_0) + \delta \Big]$ and $E_L$ = $t_2 \Big[1 + \delta \cos (2k_0)\Big]$, producing a constant chiral chemical potential $\mu_{ch}$ = $(E_R - E_L)/2$ = $t_2(\delta - 1) \sin^2 k_0$, which is essential to obtain a non-zero CPG response. One can get type-I and type-II WSM by tuning the ratio of $\frac{t_2}{t_1}$. For $\frac{t_2}{t_1}<0.01$, two bands meet at four type-I WNs. For $\frac{t_2}{t_1}>0.01$, the WNs start to tilt in the $x$-direction and we have four type-II WNs. 
Considering $k_0=\pi/4$, one finds two left chiral WNs at ${\bm k}_-^{1,2}=\pm (\pi/4 , -\pi/4 , 0)$ and 
two right chiral WNs at ${\bm k}_+^{1,2}=\pm (\pi/4 , \pi/4 , 0)$ with energies $E_L (E_R)=t_2  (t_2\delta)$.


The low energy Hamiltonian close to a given chiral node with chirality $s$ is given by ${\mathcal H^{II}}_{\bm k,s} \approx n_{s,0}\sigma_0 + t_1 ( s k_y   + \delta k_z^2/2 ) \sigma_x + t_1 k_z \sigma_y + t_1( s k_x  + \delta k_z^2/2 ) \sigma_z$ with $n_{s=-1,0}=t_2 \delta (k_y -k_x) + t_2 (1-k_x k_y)$ and $n_{s=+1,0}=t_2 \delta (1+k_x k_y) - t_2 (k_x + k_y)$. For simplicity we consider $t_1=1$.  One can now obtain the Berry curvature and the velocity difference around the right chiral WNs as $\Omega_i=\pm k_i/ k^3$ and $\Delta v_i=k_i/ k$. Following the same line argument as presented for TRS broken WSM, we find CPGE  will be governed by the two right (left) chiral WNs when $\mu \approx E_R ~(E_L)$. To be precise, quantization would be twice of the topological charge associated with the individual WNs as the contribution for two WNs with same chirality gets added up. Therefore, low energy model suggests that the CPGE (\ref{CPG_TRS})
for TRS  invariant case become twice of that of the for TRS broken case.  
 Based on the above argument considering the  low energy model, the CPGE is expected to show  quantization at two exactly opposite values irrespective of the details of the  lattice model. We shall investigate the validity of this expectation extensively in Sec.~\ref{sec3} by examining the lattice  models (\ref{eq:ham1}) and (\ref{eq:ham2}).


 For completeness, the explicit expressions of the Berry curvature ${\bm \Omega}({\bm k})=(\Omega_x(\bm k),\Omega_y (\bm k),\Omega_z(\bm k))$ and the velocity ${\bm v}(\bm k)=(v_x(\bm k),v_y(\bm k),v_z(\bm k))$ associated with the Hamiltonian (\ref{eq:ham2}) is given by \begin{eqnarray}
 \Omega_x &=& \pm \frac{(-\delta + (\cos k_0 + \delta - \cos k_x) \cos k_z) \sin k_y }{|N_{\bm k}|^3} \nonumber \\
 \Omega_y &=& \pm \frac{(-\delta + (\cos k_0 + \delta - \cos k_y) \cos k_z) \sin k_x }{|N_{\bm k}|^3} \nonumber \\
 \Omega_z &=& \pm  \frac{\sin k_x \sin k_y \sin k_z}{|N_{\bm k}|^3}\nonumber \\
 v_x &=& \pm \frac{(\cos k_0 - \cos k_x + \delta (1-\cos k_z)) \sin k_x }{|N_{\bm k}|} \nonumber \\
 &-& t_2 \delta \sin(k_x-k_y) -t_2 \sin(k_x +k_y) \nonumber \\ 
 v_y&=& \pm \frac{(\cos k_0 - \cos k_y + \delta (1-\cos k_z)) \sin k_y }{|N_{\bm k}|} \nonumber \\
 &+& t_2 \delta \sin(k_x-k_y) -t_2 \sin(k_x +k_y) \nonumber \\
v_z &=& \pm 
\frac{(2\delta^2 + \cos k_z) \sin k_z}{|N_{\bm k}|} \nonumber \\
&\pm & \frac{\delta (2 \cos k_0 - 2 \cos k_z -\cos k_x-\cos k_y) ) \sin k_z}{|N_{\bm k}|}\nonumber\\
\end{eqnarray}
For a TRS invariant WSM, we here find ${\bm \Omega} (\bm k)= -{\bm \Omega} (-\bm k)$. Here, $\pm $ refers to the valence and conduction band. The structure of Berry curvature and velocity over the lattice BZ can not be fully captured by the low energy model.  Therefore,
we shall below study the lattice model to get more reliable understanding that could relate to the experimental findings.


\section{Result and discussions}
\label{sec3}

\begin{figure*} [ht] 
\centering
\includegraphics[scale=0.2]{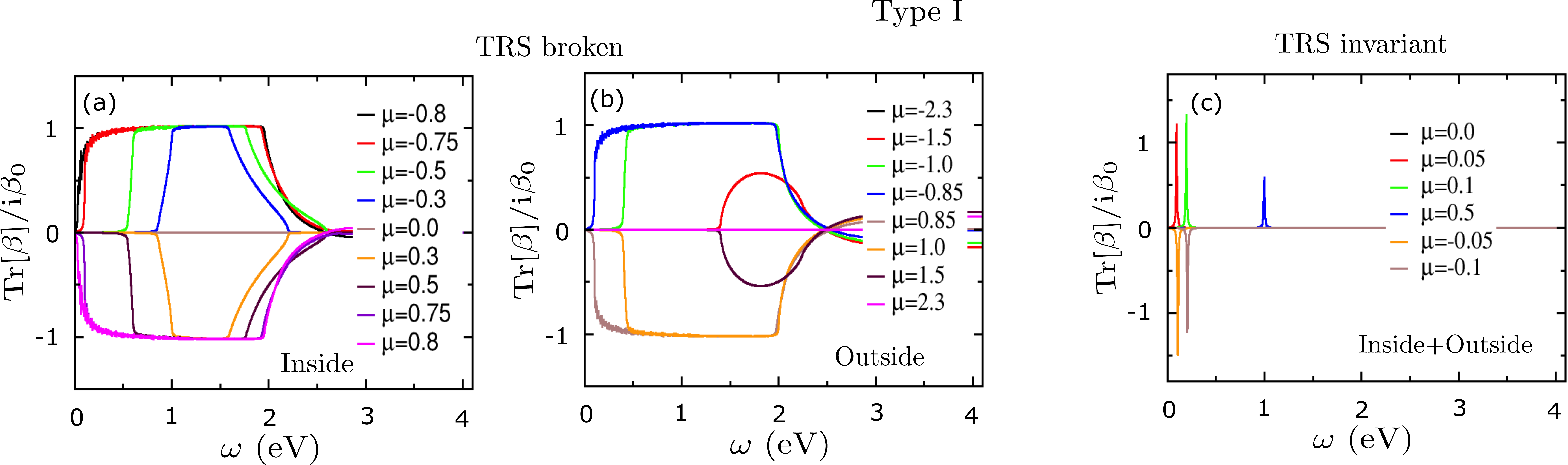} 
\caption{ Beheavior of CPG trace for TRS broken type-I WSM in (a) and (b); IS broken type-I WSM in (c) for both inside ($E_L < \mu < E_R $) and outside ($ \mu < E_L,   \mu > E_R $) region. For the TRS broken case,
the quantization in CPGE to two opposite values, that are given by the topological charge of the activated WN,  is clearly observed when $\mu$ is set in between the two WN energies. Interestingly, when $E'_{R} = E'_{L}$, CPG response appears to vanish; 
CPGE shows non-quantized behavior when $\mu$ is well above and below from WNs energies ($E_{R,L}$). In contrary, for the IS broken case, CPGE is never found to be quantized in any of the above circumstances. We consider ${\bm k}$ mesh size $(500)^3$ for our numerical computations.
}
\label{fig:comparison_CPG_typeI} 
\end{figure*}

Having discussed the formalism to compute the CPG tensor, we now investigate it for IS broken WSMs. To begin with,  we numerically estimate CPG trace for the TRS broken type I Weyl semimetal (\ref{eq:ham1}) as shown in 
Fig. \ref{fig:comparison_CPG_typeI} (a)-(b). 
We here consider the chemical potential for both inside and outside region of two non-degenerate WNs. The WNs with topological charge $C_{R,L}=\mp 1$
appear at $E_{R,L}=\pm 0.8$ for $k_0=\pm \pi/2$.  Both the WNs with energies $E_L$ and $E_R$ are equally spaced  below and above for the chemical potential $\mu = 0$ i.e., $E'_{R} = E'_{L}$ with
$|E_{R,L}-\mu|=E'_{R,L}$. We find that CPG trace vanishes irrespective of the value of frequency for $\mu = 0$. On the other hand, for $\mu=\pm 0.8$, our investigation shows that the  quantization in CPG trace at $\mp 1$ starts from $\omega=0$ and lasts until $\omega \approx 2.0$.  
However, the CPG trace decreases for $\omega >2.0$ and vanishes around $\omega\approx 3.2$. 
One can thus infer that CPG response is dependent on $\vert E'_{R} - E'_{L} \vert$. Precisely, the region of the quantization is found  inside the following frequency window $2\vert E'_{L}\vert <  \omega  < 2\vert E'_{R}\vert$.

\par We shall now discuss the CPG response when 
$\mu$ is away from the WN energies.
For the chemical potential $\mu=-0.3$, inside between two WNs with $\vert E'_{L} \vert < \vert E'_{R} \vert$, the frequency window for the quantization at $+1$  is $ 1.0 < \omega <  1.5$ (see Fig. \ref{fig:comparison_CPG_typeI} (a)). For  $\mu=0.3$, the value of the quantization reverses within the same energy windows as $ \vert E'_{L} \vert > \vert E'_{R} \vert$. The underlying reason is that the transport is maximally governed by the nature of the activated WN i.e., the magnitude (sign) of quantization depends on the topological charge (chirality) of that WN. However,  we find that  the CPG trace becomes finite
within the frequency window $2\vert E'_{L}\vert <  \omega  < 2\vert E'_{R}\vert$. The frequency above (below) which CPG trace starts (ends) showing quantized behavior decreases 
toward zero when $E'_R$ and $E'_L$ are maximally deviated from each other. As a result, for $\mu=\pm 0.8$ $(\pm0.3)$, one can find largest (smallest) frequency window for quantization. 
 When the chemical potential is outside the energy window between the two WNs, but close to any of the WNs within the linear band touching region, the CPGE is also found to be quantized (see Fig. \ref{fig:comparison_CPG_typeI} (b)). Expectedly, the quantized value depends on the topological charge of the activated WN. But when $\mu \gg E_R, E_L$, i.e., far away from the non trivial band crossing, the CPG trace becomes non-quantized acquiring  smaller value $<1$. 
Importantly, we find anti-symmetric behavior  of CPG response symmetrically placed around $\mu=(E_L +E_R)/2$.

\par In contrary, for the IS broken case of type-I as shown in Fig.~\ref{fig:comparison_CPG_typeI} (c), CPG trace is never found to be quantized within an extended window of $\omega$
in any of the above circumstances. One can observe sharp peak for certain values of chemical potential otherwise, it remains zero throughout the whole frequency range.
It can acquire values such as  $>+1$ ($<-1$) which is larger (smaller) than the topological charge of a single WN. We note that the
total number of Weyl points present in the system is four. Therefore, non-quantized CPG trace can be in principle  larger (smaller) than $+1$ ($-1$).
Based on our analysis in these two  species of type-I models, CPG trace is able  to capture the symmetry mediated transport in a distinct way. For TRS broken model, the quantized value is proportional to the charge of the WN while for TRS invariant model, the quantization is absolutely absent, however, the magnitude can be larger than the topological charge. 
One can infer that in order to obtain quantized response of CPG trace for type-I WSMs, the breaking of TRS plays a very crucial role. However, the most essential condition to obtain quantized response is to have a substantial energy gap between WNs of different chiralities.  For TRS invariant model (\ref{eq:ham2}), the above criterion 
is violated [($ E_{R}  -  E_{L} )= t_2(\delta-1) = 0.002$ eV]
while TRS broken model (\ref{eq:ham1}), it is satisfied [($ E_{R}  -  E_{L} )= 2 \gamma = 1.6$ eV]. 
In order to understand this phenomena in more detail, we below investigate the type-II analogue of these models.

{We shall now try to anchor the above numerical findings with plausible physical understanding. We first make resort to  the CPGE  formula given in Eq.~\ref{beta} where the expression inside the ${\bm k}$-sum can be decomposed in two parts namely,
 $\Delta f_{\bm k,12}  \Delta v_{i,12}\Omega_{i,\bm k}$ that does not depend on $\omega$ and the remaining part 
$ \delta(\hbar\omega - E_{\bm k,12})$ 	that only depends on $\omega$.
The CPGE obtained in Fig. ~\ref{fig:comparison_CPG_typeI} (c) clearly refers to the
fact that response is dominated by $\delta$-function as it acquires finite value only within a very short interval of $\omega$. A close inspection of the Fig.~\ref{fig:comparison_CPG_typeI} (c) suggests that 
the CPGE only becomes finite for $\omega \simeq 2 \mu +  O(t_2)$. The CPGE is expected to show finite response $2|E'_L|<\omega < 2 |E'_R|$ with $E'_{L,R}=E_{L,R}-\mu$ \cite{de2017quantized}. 
Now for the present TRS invariant type-I WSM, $E_L$ and $E_R$ both become vanishingly small with $t_2 \to 0$. The frequency interval thus shrinks to $2 (\mu - \eta_1) < \omega < 2 (\mu - \eta_2)$ with $\eta_1=t_2 \delta$ and $\eta_2=t_2$ and $\eta_{1,2} \to 0$ as $t_2 \to 0$.  Hence the part 
$ \delta(\hbar\omega - E_{\bm k,12})$  only feeds the part $\Delta f_{\bm k,12}  \Delta v_{i,12}\Omega_{i,\bm k}$  in the  momentum integral for $\omega$ centered around $\omega \simeq 2 \mu - (\eta_1 +\eta_2)$ while computing $\rm Tr[\beta(\omega)]$. For the type-I case with $t_2/t_1 < 0.01$ and $\mu > t_2$, the peak position of CPGE depends on $\mu$ for fixed value of $t_2$ and $\delta$ while these two parameters determine the shift in the peak location from $\omega = 2 \mu$.    These features are clearly noticed in Fig. ~\ref{fig:comparison_CPG_typeI} (c).}

{On the other hand, since the energies $E_L$ and $E_R$ of two WNs are substantially close to each other, optically-activated momentum surfaces, determined by the factor $\Delta f_{\bm k,12}  \delta(\hbar\omega - E_{\bm k,12})$, might embed the  non-linear band crossings. This can cause an apparent deviation from the quantization if there is any such extended region on $\omega$ within which $ \delta(\hbar\omega - E_{\bm k,12})$ acquires finite value. The magnitude of peak decreases when $\mu$ is substantially away from $E_L$, $E_R$ might refer to the fact that the contribution coming from optically activated momentum surface reduces. 
We note here that CPGE is found to be quantized in several WSMs and multifold fermionic systems with considerably separated WNs in energy space using first principle studies  \cite{flicker2018chiral,zhang2018photogalvanic,ccle20}. The non-quantized behavior of CPGE, found in Fig. ~\ref{fig:comparison_CPG_typeI} (c), is not expected to persist when $\mu_{ch}$ becomes finite. We believe that our findings on TRS invariant type-I WSM is not universal for any such pairs of WNs that are sufficiently separated in energy space. 
However, for TRS invariant WSM lattice models with WNs at different energies, the CPGE has not been studied so far due to lack of such lattice models in literature \cite{trivedi17,RevModPhys.90.015001}. The conventional expectation of quantized CPGE is solely based on the finite nature of $\mu_{ch}$ that in our case no longer holds resulting in such unconventional $\delta$-function like response. 
}

\begin{figure} [ht] 
\centering
\includegraphics[scale=0.15]{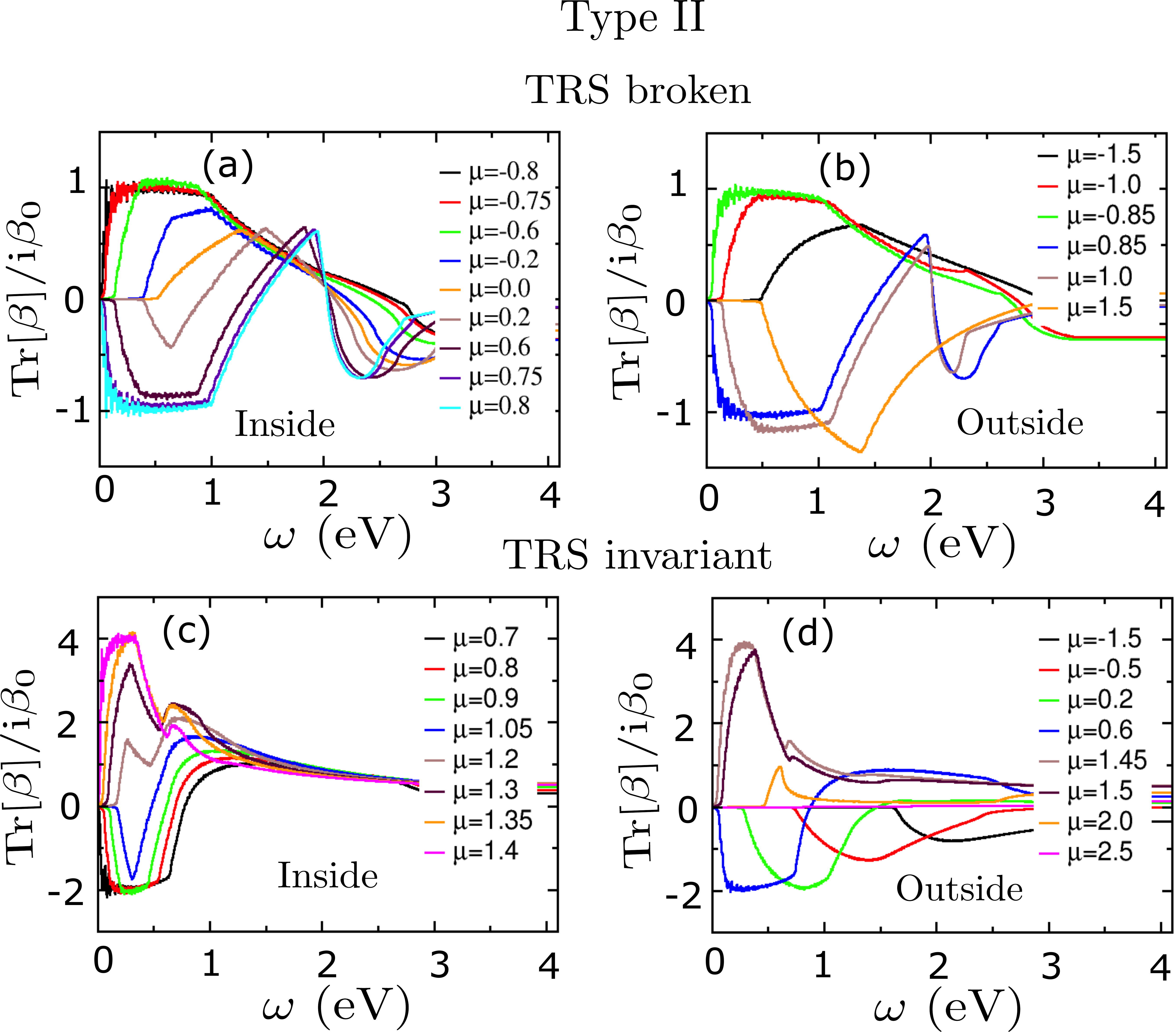} 
\caption{ Beheavior of CPG trace for TRS broken type-II WSM in (a) and (b); IS broken type-II WSM in (c) and (d) for both inside ($E_L < \mu < E_R $) and outside ($ \mu < E_L,   \mu > E_R $) region. For TRS broken case, the quantization in CPGE to two opposite values, that are given by the topological charge of the activated WN, 
is clearly observed in when $\mu$ is set only around any of the Weyl point energy ($E_{R,L}$). Interestingly, when $E'_{R} = E'_{L}$, CPGE does not to vanish like type I. For the IS broken case, the quantized value is noticed to be 2 times and  4 times the topological charge of the activated  WNs
when $\mu$ is set around $E_{L}$, $E_{R}$ respectively. We consider ${\bm k}$ mesh size $(500)^3$ for our numerical computations.}
\label{fig:comparison_CPG_typeII}
\end{figure}

\par Figure \ref{fig:comparison_CPG_typeII} (a)-(b) show the CPG trace for IS and TRS broken tilted type-II  WSM. Here the quantization is only obtained when the chemical potential is kept near to the energy of one of the WNs. For $\mu = \pm 0.8$ and $\pm 0.75$, the CPG trace is quatized with values $\mp 1$ within the frequency windows $0.2 <  \omega <  1.2$. This quantization window for type-II WSM is almost half as compared to that of the for type-I WSM with the same value of chemical potential.   The tilt modifies the available states near the Fermi surface (otherwise point like for type-I untilted case) appearing in the CPG trace  through the Fermi distribution function ( $\Delta f_{12}$) associated with the $\bm k$ modes in BZ.  The tilt thus imprints its effect by eventually normalizing the  frequecy window within which CPG response acquires quantized value. Interestingly, when $E'_L=E'_R$ for $\mu=0$, CPG trace does not to vanish like type-I rather it shows non-quantized behavior.

When $E_L<\mu<E_R$ is well separated from the Weyl point energies $E_{R,L}=\pm 0.8$, the CPG trace becomes non-quantized. Similar to the type-I WSM, we find  that
CPG trace is also quantized even when $\mu$ is kept outside the energy window between WNs but close to one of the WNs as shown in Fig. \ref{fig:comparison_CPG_typeII} (b). However, for $\mu\gg |E_{R,L}|$, one can obtain non-quantized value of CPG trace that is larger or smaller in magnitude than the absolute value of the topological charge of the activated WN. 
Moreover, the anti-symmetric nature of CPG response is not observed for type-II. These features in type-II TRS broken WSM are in stark contrast to the type-I counterpart of the same model. 
As discussed above  that the Fermi surface states contribute to the transport, any change in Fermi  surface character would be  clearly visible in the CPG response for type-I and type-II TRS broken  WSM.

Now, we analyze the CPG response for IS broken type-II WSM where we  find the quantized response for  $E_L<\mu<E_R$ kept close to the WN energy 
$E_L=t_2=0.7$ and $E_R=t_2\delta=1.4$ (see Fig. \ref{fig:comparison_CPG_typeII} (c)). This behavior remain unaltered when $\mu$ is
close to $E_L$ or $E_R$ but 
outside the  energy window set by these energies (Fig. \ref{fig:comparison_CPG_typeII} (d)). Comparing with type-II TRS broken WSM, we find that TRS invariant type-II WSM behaves in an identical way as far as the quantization is concerned.  The frequency window for
quantized response of CPGE in the TRS invariant case with $\mu \approx E_L$ is  larger than that of the for $\mu \approx E_R$. In the case of TRS broken, these 
two quantization windows appear to be similar. 
Surprisingly, CPGE becomes quantized to two different values $-2$ and $+4$ for $\mu$ close to $E_L$ and $E_R$, respectively. This suggests that the anti-symmetric nature of the CPG response is lost considering $\mu$ being symmetrically placed around $(E_L+E_R)/2$. 
This is in complete contrast to the TRS broken case where  the magnitude of quantized value depends only on the charge of the activated Weyl point.  One can find that there exist two left (right) chiral Weyl points at $E_L$ ($E_R$) with topological charge $C_L=-1$ ($C_R=+1$). When $\mu$ is set close to $E_L$, the transport is governed by both of these two left chiral WNs and they contribute additively resulting in 
CPGE to be proportional to $2 C_L$. On the other hand, when $\mu$ is close to $E_R$ where there exist two right chiral WNs with $C_R=+1$, CPGE is found to be quantized at $4C_R$ instead of $2C_R$.  This can be understood in the following way that activated WNs contribute differently i.e., the product of the Berry curvature and velocity difference in the CPG trace at left and right chiral WNs are not identical for TRS invariant model. Whereas, for TRS broken model, the product of the Berry curvature and velocity difference in behave in an identical fashion around two opposite chiral WNs which leads to 
perfectly anti-symmetric nature of CPG trace. 
 We can thus comment that the transport in type-II TRS broken WSM  is intrinsically different from the TRS invariant model type-II WSM.

\begin{figure} [b!] 
\centering
\includegraphics[scale=0.32]{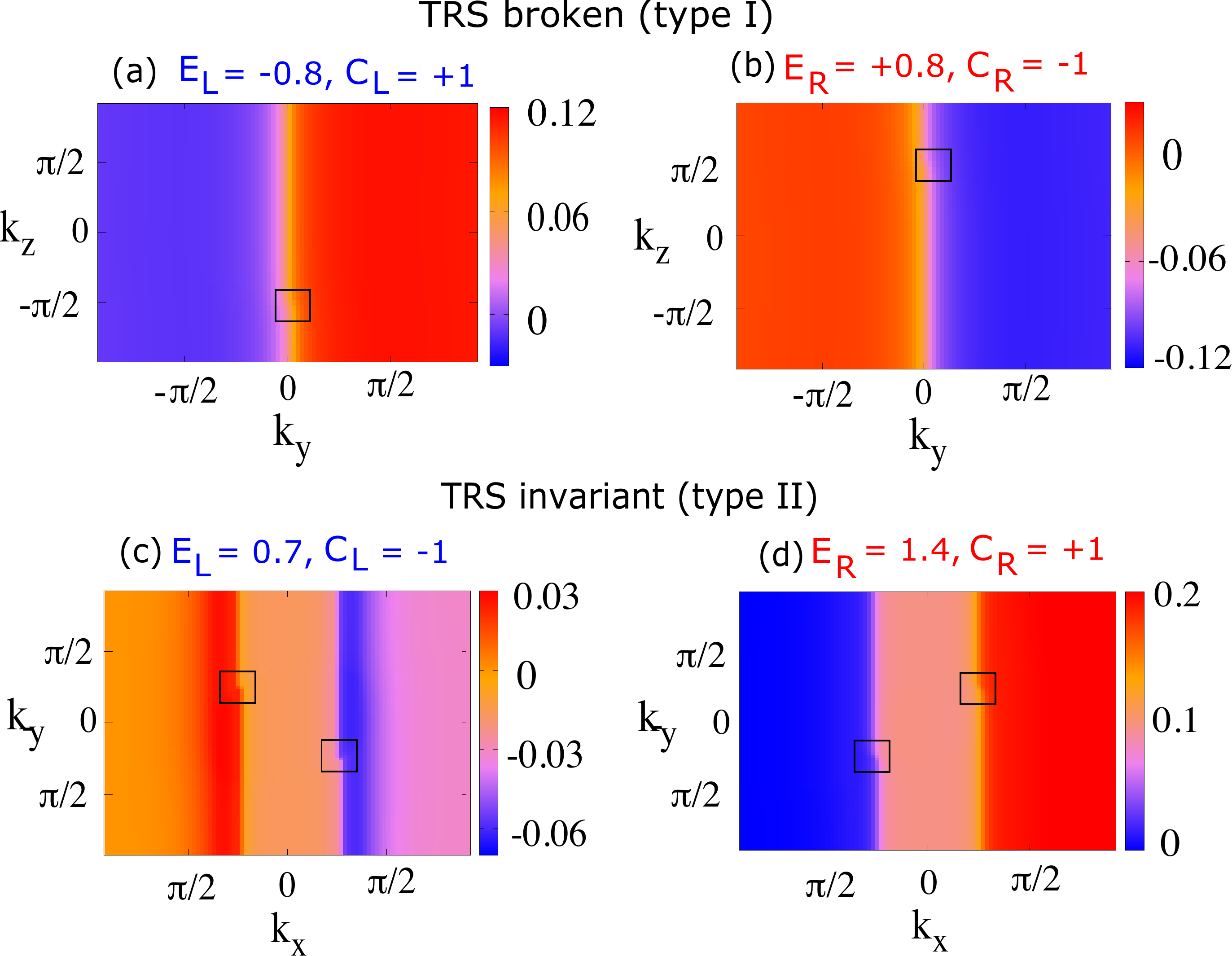}
\caption{ The $k$-resolved plots $A(\bm k, \mu)$, (a)-(b)  for TRS broken Hamiltonian ${\mathcal H}^{I}({\bm k})$, and (c)-(d) for TRS invariant Hamiltonian ${\mathcal H}^{II}({\bm k})$ with  $\mu=E_L, E_R$ respectively. Here, we consider  $t_2=0$ and $\gamma=0.8$ in ${\mathcal H}^{I}({\bm k})$;
$t_2=0.7$ and $\delta=2.0$ in ${\mathcal H }^{II}({\bm k})$. 
The position of the WNs are marked by square. 
Comparing the momentum structure of $A$, one can infer that CPGE at $\mu=E_L$ acquires opposite values to that of at $\mu=E_R$ in (a) and (b) for TRS broken WSMs, whereas, CPGE acquires opposite but non anti-symmetric values at $\mu=E_L$ and $\mu=E_R$ in (c) and (d) for TRS invariant WSMs.}
\label{fig:k_resolved}
\end{figure}

We would now like to understand our results more deeply from the physical point of view.  Using the CPG tensor (\ref{beta}), one can find for the TRS broken WSM that ${\rm Tr}[\beta] \simeq  \sum_{\bm k} f_{\bm k,12} \sum^{x,y,z}_i \Delta v_{i,12}(\bm k)\Omega_i(\bm k) \delta(\omega - E_{\bm k,12} ) $ with $A(\bm k, \mu)=f_{\bm k,12} \sum^{x,y,z}_{i}\Delta v_{i,12}(\bm k)\Omega_i(\bm k)$. It is better to focus 
on the CPG response when $\mu$ is chosen 
close to the WNs energies as the transport is maximally controlled by the nature of the  WNs. For TRS broken WSM, we can infer that   $A(\bm k_s, \mu \approx E_{R,L}) \simeq s \sum^{x,y,z}_i \Delta v_{i,\bm k_s}\Omega_{i,\bm k_s} $  with $s=\mp 1$ substantially dominates in determining the behavior of ${\rm Tr}[\beta]$. 
 A close inspection, considering the low energy model, suggests that  $A(\bm k_-, \mu \approx  E_{R}) = -  A(\bm k_+, \mu \approx E_{L})$ as ${\bm \Omega}(\bm k_+)= {\bm \Omega} (\bm k_-)$ and $ \Delta {\bm v} (\bm k_+)= - \Delta {\bm v} (\bm k_-)$. This  leads to the fact that the injection current changes its sign as $\mu$ switches from left chiral WN energy to right chiral WN energy. In order to anchor this analytical analysis, we study 
 $A(k_x=0,k_y,k_z,\mu)$ numerically from the lattice model (\ref{eq:ham1}) in Fig.~\ref{fig:k_resolved} (a) and (b) for $\mu=-0.8$ and $0.8$, respectively. We find that the sign of $A$ reverses for $\mu=\pm 0.8$ with $k_y >0$. The WN at ${\bm k}_+=(0,0,-\pi/2)$ for $\mu=-0.8$ actively participates in CPGE quantization with positive magnitude in the sense that a kink is observed in  $A$; the same observation but negative in magnitude is also noticed for the  WN at ${\bm k}_-=(0,0,\pi/2)$ with $\mu=0.8$.

  We would now analyze the TRS invariant case where four Weyl points are found; two left chiral WNs at ${\bm k}^{1,2}_-=\pm (\pi/4,-\pi/4,0)$ with $E_L=t_2$ and two right chiral WNs at ${\bm k}^{1,2}_+=\pm (\pi/4,\pi/4,0)$ with $E_R=t_2 \delta$.
 When $\mu \approx  E_L$, the left chiral nodes contribute maximally  to the CPG tensor  ${\rm Tr}[\beta] \simeq \sum_{\bm k}  \sum^{1,2}_i A({\bm k}^i_-, \mu  \approx E_L) \delta(\omega - E_{\bm k,12} )$.  The analysis from low energy model refers to the fact that CPGE acquires $2si\beta_0$ as the contribution from two WNs with same chirality gets added up. Therefore, the non anti-symmetric behavior of CPGE trace as observed in \ref{fig:comparison_CPG_typeII} (c)-(d) can not be explained by the low energy model. The structure of $\Omega(\bm k)=-\Omega(-\bm k)$, as observed in TRS invariant WSMs, can not be captured in the corresponding low energy model. The same applies for the velocity difference  $\Delta v$ also. These can result in a distinct behavior as compared to the low energy model while a TRS invariant lattice model is considered. Interestingly, in the TRS broken case all three component of CPG tensor i.e., $\beta_{xx}$, $\beta_{yy}$ and $\beta_{zz}$, contribute equally i.e., 
 $\beta_{xx}=\beta_{yy}=\beta_{zz}$,
irrespective of the fact that whether $\mu \approx E_L$ or $E_R$. For TRS invariant WSM, this analogy breaks
$\beta_{xx}=\beta_{yy}=\beta_{zz}/2=i\beta$. More interestingly, 
the magnitude of individual component $|i\beta|$, evaluated for $\mu \approx  E_L$,  gets doubled   $2 |i\beta|$ while computed for $\mu \approx  E_R$. 
 This causes the sharp contrast to the TRS broken case where the magnitude of quantization for CPG response becomes identical for both the chemical potential $\mu  \approx E_L$ and $\mu \approx E_R$.
Furthermore, the frequency window for quantized CPGE in this case becomes different for $\mu \approx  E_R$ and $E_L$. By contrast, this frequency window  for TRS broken case is similar.

  In order to anchor this analysis, we show  $A(k_x,k_y,k_z=0,\mu)$ numerically from the lattice model (\ref{eq:ham2}) in Fig.~\ref{fig:k_resolved} (c) and (d) for $\mu=E_L=0.7$ and $\mu=E_R=1.4$, respectively.  We find that the sign of $A$ reverses between $k_x >\pi/4$ and $k_x <-\pi/4$ with $\mu=0.7$. While for $-\pi/4 < k_x < \pi/4$, $A(k_x,k_y,k_z=0,\mu=E_L) \sim 0 $. $A$ exhibits  kink close to 
  the Weyl point ${\bm k}^{1,2}_-=\pm (\pi/4,-\pi/4,0)$ for $\mu=0.7$ rendering the fact that these two left chiral nodes actively participate in CPGE quantization which is found to be $-2$. 
  While for $\mu=1.4$, $A>0 (\sim 0)$ for $k_x> -\pi/4 ~(< -\pi/4)$; however, the value of $A$ increases for  $k_x> \pi/4$ as compared to the value of $A$ for $-\pi/4 < k_x < \pi/4$. 
  Here, $A$ exhibits  kinks close to the right chiral Weyl point ${\bm k}^{1,2}_+=\pm(\pi/4,\pi/4,0)$ suggesting the fact that these WNs actively participate in the quantization in CPG response which has the value  $\\
  4$.

The momentum integration of $A$ over the BZ for TRS broken WSM becomes negative ( positive) when $\mu=0.8$ ($\mu=-0.8$) as shown in Fig.~\ref{fig:k_resolved} (a) (Fig.~\ref{fig:k_resolved} (b)). This is directly reflected in the behavior of CPG tensor for $\mu=\pm 0.8$. Once $\mu$ reduces (increases) from $0.8$ ($-0.8$), the magnitude of $A$ decreases in the BZ for  $k_y>0$. For $\mu=0$, $A$ becomes vanishingly small in the BZ. As a result, the quantization is observed for $-0.8\le \mu \le0.8$ except $\mu=0$. These above nature of the momentum distribution of $A$ is also qualitatively valid for the tilted type-II case.  However, unlike the type-I case, the tilt can destroy the quantization when $\mu$ is away from $\pm 0.8$ within the window $-0.8 \le \mu \le 0.8$. The non-zero value of CPG tensor for $\mu=0$ in type-II case is due to the anisotropic nature of the dispersion which is imprinted in $A$ through the Fermi distribution function $\Delta f_{12}$.

The lattice analysis of $A$  further reveals that it can have both positive and negative contributions in the BZ for TRS invariant WSM with $\mu=0.7$ as shown in Fig.~\ref{fig:k_resolved} (c). Upon increasing $\mu>0.7$, one can find that $A$ reduces for $k_x<-\pi/4$.  $A$ increases and becomes positive when $\mu$ reaches $1.4$ when $k_x<-\pi/4$. In intermediate zone $-\pi/4 < k_x < \pi/4$, $A$ increases when $\mu$ increases from $0.4$ to $1.4$.  
For $\mu=1.4$, $A$ only acquires positive values for $k_x>\pi/4$ as shown in Fig.~\ref{fig:k_resolved} (d). 
From the momentum distribution of $A$ for the TRS invariant WSM, it is evident that   
$A$ does not show any anti-symmetric behavior as observed for TRS broken WSM when $\mu$ is kept at two different Weyl point energies $E_L$ and $E_R$. This again points towards the fact that CPG response can be very different for TRS broken and TRS invariant WSM in terms of the quantization. On the other hand, for type-I TRS invariant WSM, $A$ turns out to be vanishingly small but anti-symmetric with respect to $k_x=0$-plane in the BZ. As a result the CPG response becomes characteristically different from the tilted case.

\par Exact quantization in CPG trace is predicted from the ${\bm k}\cdot{\bm \sigma} $ model. The quantization can be  destroyed due to several lattice effect. Away from the WNs, band bending causes the deviation of CPG response from quantized value. This quantization is clearly observed when $\mu$ is set close the WN energy. On a general note, we can infer that for type-II WSM, the band bending near the WNs affects the quantization more compared to type-I. The band bending is more prominent in type-II and that can result in the non-linear correction to the quantization. 
In addition, we would like to note   for experimental observability that the prefactor $\beta_0$ of
Eq.~(\ref{beta}) is large in comparison to ordinary CPG trace magnitudes.
Considering typical relaxation times, one can find that the
 other metallic or insulating contributions are
less than an order of magnitude as compared to the
quantized WN contribution \cite{moore2010confinement, sodemann2015quantum}. As a result, we believe total CPG trace observed in experiment can signal the quantization \cite{ni2020giant}.

{ We note that due to the lack availability of TRS invariant WSM lattice model, hosting WNs of opposite chirality at two different energies, we could not generalize our findings in other TRS invariant WSM models with $\mu_{ch} \ne 0$ \cite{trivedi17,RevModPhys.90.015001,computation}. Interestingly, $\mu_{ch}$ changes with tilt parameter for our present model (\ref{eq:ham2}) that further causes the CPGE to behave distinctly in type-I and type-II phases (see Fig.~\ref{fig:comparison_CPG_typeI} and Fig.~\ref{fig:comparison_CPG_typeII}). In general, due to the change in Fermi surface properties, type-II WSMs exhibit many intriguing transport signatures \cite{PhysRevLett.117.077202}. 
On the other hand, the low energy model, generically defined for a single WN, might not distinguish a TRS invariant WSM from a TRS  broken WSM. Our study suggests that the CPGE for TRS invariant WSM can not be thoroughly explained by the low energy model contrasting the CPG response for
TRS broken WSM. This refers to the fact that lattice model is indispensable in the present case of TRS invariant WSM \cite{PhysRevB.94.245143}.     
Unlike the TRS broken WSM hosting a single Fermi arc of definite chirality, TRS invariant WSM supports a  pair  of chiral  Fermi  arcs.  This has a severe implication in transport properties such as dc and optical conductivities  \cite{PhysRevB.97.165201,PhysRevB.93.085426}. We emphasize that our work thus uncovers various aspects of the transport signatures while investigating appropriate lattice models for tilted WSMs in presence as well as absence of TRS. Therefore, our findings indeed convey a 
general picture, that is not limited to the specific models considered here, as the conventional expectations are equally valid and are emerged from the same physical origins. }

\section{conclusion}
\label{conclusion}

We consider TRS invariant and TRS broken WSM to study the CPG response where the energies $E_L$ and $E_R$, associated with the left and right chiral Weyl points,  are different from each other  $E_L \ne E_R$. The motivation is to analyze the effect of TRS on the quantization while the IS and mirror symmetries are already broken for both these WSMs. We consider general tilted lattice Hamiltonian which allows us to additionally investigate the effect of tilted dispersion in the CPG response. There exist only two WNs of opposite chirality in TRS broken system, while at least four WNs of opposite chiralities are present in TRS invariant WSM. Therefore, when the number of WNs for a given chirality is more than unity that could become an interesting situation to study. To be precise, a relevant question here is that how does the quantization depend on the number of WNs.
In this work, we show that quantization for TRS invariant single WSM can be $2$ and $4$ times the topological charge of the activated WNs (see Fig.~\ref{fig:comparison_CPG_typeII}). This feature is not observed for TRS broken WSM as there exists only one WN for a given chirality (see Fig.~\ref{fig:comparison_CPG_typeI}). In addition, CPG response is able to distinguish a type-II from a type-I WSM in general.
{It is noteworthy that  non-quantized peak in CPGE at certain frequecy,
depending on the values of chemical potential, are originated due to fact that $E_L \simeq E_R$ for TRS invariant type-I WSM. This is in stark contrast to all other quantized CPG responses where $E_L$ and $E_R$ are substantially separated from each other. }

In particular, the Berry curvature and the velocity difference play very important role in determining the behavior of CPG trace. The tilt is not able to change the Berry curvature for anisotropic case from the isotropic case, however, velocity along the tilt direction can become very different from the isotropic case.  The effect of tilt enters through the Fermi distribution function in the CPG trace as the velocity difference remains independent of the tilt parameter.
As a result, CPG response for type-II is distinguishably different from type-I. For example, in the TRS broken type-I case, CPG trace behaves exactly opposite to each other when $\mu$ is symmetrically chosen around $\mu= (E_L + E_R)/2$. This feature is not observed for the tilted type-II case. Moreover, CPG response can acquire values larger than the magnitude of the topological charge in presence of the tilting. 
Interestingly, in our present case, type-II TRS invariant can only exhibit quantization in CPG trace unlike to the type-I counterpart where $E_L \approx E_R$.
The magnitude of quantization for $\mu$ close to $E_L$ and $E_R$ strongly depends on Berry curvature and velocity around the left and right chiral WNs. In the TRS broken WSM, the value of quantization  for $\mu \sim E_L$ is just opposite to that of the for $\mu\sim E_R$. For TRS invariant these two values of quantization are different from each other as the Berry curvature and velocity (as well as the velocity difference) behave differently for left and right chiral WNs. We can comment that low energy model might not always predict the numerical findings, based on the lattice model, specially when there exists more that a single WN with a given chirality \cite{computation}. This is the case we encounter in TRS invariant WSM lattice Hamiltonian and the non anti-symmetric quantization of CPGE there might not be explained by the associated low energy model.

We believe that our observation can be tested experimentally due to availability of the setup. It would be interesting to study the TRS broken and invariant chiral multi-WSM having non-linear anisotropic dispersion. The band bending causes the CPGE to deviate from quantized behavior. The role of TRS and band bending are two important aspects that can be studied in future in chiral SMs \cite{schroter2019chiral, changdar2020electronic,ni2020giant}.  

\section{Acknowledgements}
TN would like to thank the computation facility provided by MPIPKS, Dresden, Germany. We would like to thank Adolfo G. Grushin for carefully reading the manuscript and giving his valuable comments.

\bibliography{cpge}

\end{document}